\documentclass[preprint,5p,twocolumn]{elsarticle}

\usepackage{mathrsfs,amsmath,amsthm,amssymb,braket,color,verbatim,adjustbox,multirow,enumerate,textcomp,gensymb,subfigure,graphicx,diagbox,pifont,tabulary,booktabs,stackengine,epstopdf,dcolumn,makecell,listings,bbm,mathtools,autobreak,slashed,float,array,colordvi,xfrac}
\usepackage{lineno}
\modulolinenumbers[5]
\usepackage{hyperref}
\hypersetup{colorlinks,bookmarksopen,bookmarksnumbered,citecolor=[rgb]{0,0,0.5},linkcolor=black,urlcolor=[rgb]{0,0,0},pdfstartview=FitH,linktocpage}

\definecolor{eprintLinks}{rgb}{0,0,0.5}
\definecolor{journalLinks}{rgb}{0,0,0.5}
\newcommand{\MYhref}[3][blueLinks]{\href{#2}{\color{#1}{#3}}}

\def\D{\varDelta}
\def\O{{\cal O}}

\def\Ep{E_{\textrm{Pl}}}
\def\Elv{E_{\textrm{LV}}}
\def\Elvn{E_{\textrm{LV},n}}

\def\dla{\langle\!\langle}
\def\dra{\rangle\!\rangle}

\journal{PLB, published as Phys. Lett. B 835 (2022) 137543, \href{https://doi.org/10.1016/j.physletb.2022.137543}{doi:10.1016/j.physletb.2022.137543}}

\begin{document}
\biboptions{numbers,sort&compress}

\begin{frontmatter}

\title{Lorentz- and CPT-violating neutrinos from string/D-brane model}

\author[pku]{Chengyi Li}
\author[pku,CHEP,CICQM]{Bo-Qiang Ma\corref{em}}
\cortext[em]{Corresponding author \ead{mabq@pku.edu.cn}}

\address[pku]{School of Physics, Peking University, Beijing 100871, China}
\address[CHEP]{Center for High Energy Physics, Peking University, Beijing 100871, China}
\address[CICQM]{Collaborative Innovation Center of Quantum Matter, Beijing, China}

\begin{abstract}
We show that the space-time foam model from string/D-brane theory predicts a scenario in which neutrinos can possess linearly energy dependent speed variation, together with an asymmetry between neutrinos and antineutrinos, indicating the possibility of Lorentz and CPT symmetry violation for neutrinos. Such a scenario is supported by a phenomenological conjecture from the possible associations of IceCube ultrahigh-energy neutrino events with the gamma-ray bursts. It is also consistent with the constraints set by the energy-losing decay channels~(e.g., $e^{+}e^{-}$ pair emission, or neutrino splitting) upon superluminal neutrino velocities. We argue that the plausible violations of energy-momentum conservation during decay may be responsible for the stable propagation of these neutrinos, and hence for the evasion of relevant constraints. 
\end{abstract}


\end{frontmatter}

\section{Introduction}

As a cornerstone of modern physics, Lorentz symmetry is preeminently fundamental and profound, whereas it may be broken at very high energies~(typically approaching the Planck scale $\Ep=(\hslash c^{5}/G)^{1/2}\simeq 1.22\times 10^{19}$~GeV) due to inherent ``foamy'' structures of space-time, as implemented by various candidate models for quantum gravity~(QG)~\cite{Amelino-Camelia:1997ieq,He:2022gyk}. Generally, Lorentz symmetry violation~(LV) manifests itself via an energy dependent deviation in the velocity of an energetic photon or neutrino propagating \textit{in vacuo}, $\delta=(v-c)/c$~\cite{Amelino-Camelia:1997ieq,Jacob:2006gn,He:2022gyk}. From the phenomenological point of view, by taking into account only the leading order modification, one may parametrize $\delta$, for a particle of rest mass $m$ such as a neutrino with energy $mc^{2}\ll E\ll\Ep$, as 
\begin{equation}
\label{eq:1}
\delta\simeq -\frac{s_{n}}{2}(n+1)(E/\Elvn)^{n},
\end{equation}
where $\Elvn$ is the Lorentz violation scale~(the energy scale where QG induced $n$th-order LV effects become appreciable), and $s_{n}=+1~(-1)$, the LV sign factor, represents the ``subluminal''~($\delta<0$)~(``superluminal''~($\delta>0$)) case. The putative existence of the velocity defect~(\ref{eq:1}) then leads to a flight time difference between high-energy particles and low-energy photons. In this respect, neutrinos produced in astrophysical processes, like gamma-ray bursts~(GRBs), located at large redshifts are suggested to serve as ideal tools to test such LV effect~\cite{Jacob:2006gn}. 

The IceCube Neutrino Observatory has reported till now plenty of high-energy neutrinos probably of extragalactic origin, including also a couple of PeV events~\cite{IceCube:2013cdg,IceCube:2014jkq,IceCube:2016uab}. By using the maximum correlation criterion, Amelino-Camelia and collaborators~\cite{Amelino-Camelia:2016fuh} associate IceCube 60--500~TeV events with GRB candidates to detect LV modified laws of propagation for neutrinos in a quantum space-time. Stimulated by the study, recent works~\cite{Amelino-Camelia:2016ohi,Huang:2018ham,Huang:2019etr,Huang:2022xto} find temporal and directional coincidence of neutrinos of near-TeV to PeV energies with GRBs in an expanded time window. Analyses on the flight times of these events with respect to that of the low-energy photons of associated GRBs lead to an energy dependent neutrino speed variation\footnote{In units of conventional velocity of light $c=1$ used hereafter.}:
\begin{equation}
\label{eq:2}
v=1\mp E/\Elv\ \Leftrightarrow\ \delta=\mp\,(E/\Elv),
\end{equation}
corresponding to the case of $n=1$, $s(\equiv s_{1})=\pm 1$ of~(\ref{eq:1}), with $E_{\textrm{LV},1}\equiv\Elv=(6.4\pm 1.5)\times 10^{17}$~GeV, which is close to the Planck energy scale. Due to the fact that there could be uncertainties from the hypothesis on the associations of these events with astro-objects like GRBs when analyzing the data, we should interpret this previous result as a lower bound on neutrino Lorentz violation with 
\begin{equation}
\label{eq:3}
\Elv\gtrsim 6.4\times 10^{17}~\textrm{GeV}. 
\end{equation}
It is, nevertheless, worth noting that such a Lorentz violation bound is consistent with various time-of-flight constraints, available today, from MeV neutrinos of supernova 1987A~\cite{Ellis:2008fc} as well as that from energetic events registered at IceCube, e.g.,~\cite{Wang:2016lne}\footnote{Although the limit~\cite{Wang:2016lne} proposed therein, $\Elv>0.01\Ep$, results from an association~\cite{Kadler:2016ygj} between blazar PKS B1424-418 with a PeV event~(\#35) which is however suggested by~\cite{Huang:2018ham} to be associated with a GRB, it is still compatible with the bound~(\ref{eq:3}), as stated in~\cite{Huang:2018ham}.}; it is also compatible with the constraints~\cite{Ellis:2018ogq,Laha:2018hsh,Wei:2018ajw} from recent multi-messenger observations of blazar TXS 0506+056 coincident with $\sim 290$~TeV tracklike neutrino event~\cite{IceCube:2018dnn}.  

Interestingly, it is suggested~\cite{Huang:2018ham,Huang:2019etr,Huang:2022xto} that there exist both time ``delay''~($s=+1$) and ``advance''~($s=-1$) events. One possibility is that the associations of neutrino events with GRBs are purely accidental, while an asymmetry between neutrinos and antineutrinos~\cite{Huang:2018ham} may explain the coexistence of ``delay'' and ``advance''  events, say, either neutrinos or antineutrinos are superluminal, while the other ones are subluminal, leading to the speculation of Charge--Parity--Time~(CPT) symmetry violation~\cite{Zhang:2018otj} between neutrinos and antineutrinos. This conjecture seems very striking, so more observations and careful analyses of data are called for to check further the revealed regularity, as well as the validity of the associations of GRBs with IceCube neutrinos according to the direction and time criteria~\cite{Amelino-Camelia:2016fuh,Amelino-Camelia:2016ohi,Huang:2018ham} adopted in these works.\footnote{The correlation of IceCube events with GRBs could not be considered conclusive. The sources might have various possibilities, such as flaring blazars~\cite{Kadler:2016ygj} just mentioned, while a small number of neutrinos associated with blazars does not conflict with attributing GRB as a significant source for IceCube TeV and PeV events~\cite{Huang:2022xto}.}

Nonetheless, as we will discuss in the present Letter, this may indeed be the case in some Liouville inspired stringy analogues of space-time foam~\cite{Ellis:1992eh,Ellis:1999jf,Ellis:2000sf,Ellis:2004ay,Mavromatos:2005bu,Ellis:2008gg,Li:2009tt,Mavromatos:2012ii,Li:2021gah}, which may provide a scenario in which neutrinos possess QG induced velocity variations of the linear form~(\ref{eq:2}), due to their microscopic interactions with  membrane~(domain-wall-like) defects allowed by string/D-brane theories. The model is thus supported by the above phenomenological conjecture. In such a scenario, those severe constraints derived from the argument~\cite{Zhang:2018otj} that superluminal neutrino would in general radiate electron-positron pairs~($\nu\rightarrow\nu e^{+}e^{-}$) are evaded. It is shown that the sizable Lorentz violation for neutrinos has nothing to do with that in the charged lepton sector, where LV effect is exactly absent, for specifically stringy reasons to be outlined below, thereby escaping the constraints as developed by, e.g.,~\cite{Crivellin:2020oov}, in which the theory taken into account enjoys ordinary $\textrm{SU}(2)_{L}$ gauge invariance under the Standard Model group, which however could be broken in the stringy sense considered here. 

\section{CPT-violating neutrino from strings}

The scenario is based upon the space-time foam model of string theory~\cite{Ellis:1999jf,Ellis:2000sf,Ellis:2004ay,Mavromatos:2005bu,Ellis:2008gg,Li:2009tt,Mavromatos:2012ii,Li:2021gah}, traced back to Liouville QG~\cite{Ellis:1992eh}. In the model, our world is a~(compactified) D3-brane Universe, moving slowly in a higher-dimensional bulk space, punctured by a population of point-like D0-brane~(dubbed ``D-particle'') defects~\cite{Ellis:1999jf,Ellis:2000sf,Ellis:2004ay,Mavromatos:2005bu} which could break Poincar\'e invariance, and hence local Lorentz invariance. An intriguing feature of this scenario is that, due to unobserved~(by a low-energy braneworld observer) degrees of freedom associated with the recoil of D-particles during their interaction with the string matter resided on the brane, there could be CPT violations, in the sense of supporting asymmetric dispersion relation between neutrinos and antineutrinos, as they propagate through the ``medium'' of quantum gravity foam. 

The recoil velocity, if denoted by $u$, of a D-particle following scattering off an open-string neutrino excitation via a momentum transfer $\Delta k_{i}=\lambda k_{i}$, is $u_{i}=(g_{s}/M_{s})\Delta k_{i}$~\cite{Ellis:1999jf,Ellis:2000sf,Ellis:2004ay}, here $g_{s}\ll 1$ is the~(weak) string coupling, while $M_{s}$ is the string mass scale. Assuming that the relevant fraction ratio $\lambda$ of the incident momentum $k$ is isotropic on average, $\dla\lambda\dra=0$, but stochastic we may expect $\dla\lambda^{2}\dra=\zeta^{2}<1$, where $\dla\cdot\!\cdot\!\cdot\dra$ denotes a statistical average over collections of quantum-fluctuating D-defects encountered by the neutrino~\cite{Mavromatos:2005bu} and, as we will see, $\zeta$ is a parameter characterizing the neutrino propagation. The D-particle recoil distorts the neighboring space-time by endowing it a metric of Finsler type: $G_{\mu\nu}=\eta_{\mu\nu}+h_{\mu\nu}$, $h_{0i}=u_{i}$. This affects the energy-momentum relation~\cite{Mavromatos:2012ii} of a neutrino, of mass $m_{\nu}$, via $k^{\mu}k^{\nu}G_{\mu\nu}=-m_{\nu}^{2}$ which, upon averaging over the foam, and supposing that $k\gg m_{\nu}$, yields the average energy for neutrino species, $E\simeq\pm\sqrt{k^{2}+m_{\nu}^{2}}\bigl(1+\zeta^{2}k^{2}/(2M_{D}^{2})\bigr)$, $M_{D}=M_{s}/g_{s}$ being the D-particle mass scale. Meanwhile, the effect of defect/neutrino scattering kinematics enters the dispersion relation via $E=E^{\prime}+\dla M_{D}u^{2}/2\dra$, where $M_{s}u^{2}/(2g_{s})$ is the kinetic energy of a heavy~(nonrelativistic) D-particle and $E^{\prime}$, the outgoing average energy, yields the dispersion relation to leading order in $\zeta^{2}$~(for $\zeta^{2}\ll 1$): 
\begin{align}
\label{eq:4}
&E_{\nu}\equiv E^{\prime}=E_{M}-\zeta^{2}\frac{k^{2}}{2M_{D}},\\
\label{eq:5}
&E_{\overline{\nu}}\equiv -E^{\prime}=E_{M}+\zeta^{2}\frac{k^{2}}{2M_{D}}.
\end{align}
Here $E_{\overline{\nu}}=-E^{\prime}>0$ indicates the positive energy of a physical antiparticle~(e.g., antineutrino), and $E_{M}$ denotes the standard Minkowski energy $E_{M}\equiv\sqrt{k^{2}+m_{\nu}^{2}}$. 

Superluminality of the antineutrino is established on the assumption that the relation $v=\partial E/\partial k$ holds in QG, that is, $\delta_{\overline{\nu}}=v_{\overline{\nu}}-1=\zeta^{2}k/M_{D}>0$, while $\delta_{\nu}=-\zeta^{2}k/M_{D}$, in the case that the miniscule neutrino masses are negligible $m_{\nu}\simeq 0$~(for instance at high energies, $2k^{2}\lvert\delta\rvert\gg m_{\nu}^{2}$). Otherwise an extra term $[-(m_{\nu}/\sqrt{2}k)^{2}]$ is expected in both $\delta_{\nu}$ and $\delta_{\overline{\nu}}$~(or the velocities, $v_{\nu}$ and $v_{\overline{\nu}}$). Hence, only subluminal neutrinos are present in the theory while antineutrinos could be slightly superluminal, and neutrino propagation in vacuum is thus CPT violating. This effect scales linearly with energy, and is flavor~(i.e., neutrino-species) independent, with no effects on oscillations, therefore consists with existing bounds from neutrino oscillations~\cite{IceCube:2017qyp}. 

We observe that the above described string model is supported by the phenomenologically speculated neutrino/antineutrino propagation properties, as mentioned in the Introduction, if we recognize antineutrinos as superluminal particles while neutrinos are subluminal. This is a postulate but a reasonable one~\cite{Huang:2018ham}, because the IceCube detector does not distinguish between neutrinos and antineutrinos. The analysis result from neutrino time-of-flight studies~\cite{Amelino-Camelia:2016ohi,Huang:2018ham,Huang:2019etr,Huang:2022xto} is then viable on limiting the QG scale $M_{s}/\zeta^{2}$ for this scenario. We infer $M_{D}/\zeta^{2}\gtrsim 6.4\times 10^{17}$~GeV, or 
\begin{equation}
\label{eq:6}
\zeta^{2}<1.6\times 10^{-18}\Bigl(\frac{M_{s}}{g_{s}}\Bigr)~\textrm{GeV}^{-1},
\end{equation}
as a conservative bound on $\zeta$. Notice that the value of the mass of the foam defect, $M_{D}$, in the modern framework of the D-brane approach to QG is an adjustable free parameter. For example, $\zeta^{2}\sim\O(1)$ would imply a sub-Planckian D-particle mass $M_{s}/g_{s}\sim 10^{18}$~GeV, such that the D-foam generates a neutrino speed variation at scales $\gtrsim 10^{17}$~GeV and, thus, the CPT violation of neutrinos, compatible with the phenomenological speculation. This, in turn, serves as a support to the above model of string theory. 

It has been argued however that in cases of superluminal propagation, certain otherwise forbidden processes~\cite{Cohen:2011hx}, especially the neutral-current mediated neutrino decays by emitting $e^{+}e^{-}$ pairs, or by splitting into 3 neutrinos, are permitted, and severely deplete most of the energetic neutrinos along their path to the detectors. Such an argument, as was presented by Cohen and Glashow, was first used to refute the OPERA initial result~\cite{OPERA:2011ijq},\footnote{The OPERA ``anomaly'' was later shown to be an error resulting from a loose cable and a clock ticking too fast. The corrected result and similar terrestrial experiments~\cite{Antonello:2012be} on the speed of neutrinos all give values consistent with that of light within experimental uncertainties, but these data neither require nor exclude the possibility of superluminality~\cite{Ma:2012zd}; they were just not sensitive enough to measure any nonzero value of $\delta$ given the smallness of neutrino mass.} and was thereafter adopted as a way to impose strong new constraints upon superluminal velocities of higher-energy astrophysical neutrinos observed by IceCube. In particular, it was claimed that observations of $\gtrsim 100$~TeV diffuse neutrinos, as well as the very existence of PeV-scale neutrino events, provide bounds~\cite{Borriello:2013ala,Stecker:2013jfa,Diaz:2013wia,Wang:2020tej} as tight as $\delta<10^{-18}~\textrm{to}\sim 5.6\times 10^{-19}$ for a positive $\delta$. This corresponds to $\Elv\geq (10^{3}-10^{5})\times\Ep$ for a linear LV case~(\ref{eq:2}). 

Suppose the usual conservation laws of energy and momentum, as in Refs.~\cite{Borriello:2013ala,Stecker:2013jfa,Diaz:2013wia,Wang:2020tej} and~\cite{Cohen:2011hx}, the anomalous decay process like $\nu\rightarrow\nu e^{+}e^{-}$ does allow us to limit neutrino superluminality. However, that may not be the case in the string model discussed here, where the energy-momentum conservation may \textit{not exactly} hold due to stochastic losses $\delta E_{D}$ in particle reactions with the D-brane foam~\cite{Ellis:2000sf}. Indeed, in a four-particle interaction which may be factorized as products of three-particle vertices mediated by the exchange of an intermediate particle excitation with a momentum $p$, the energy is \textit{lost} in the stochastic D-foam medium by the amount of $(\varsigma_{I}/M_{D})p^{2}$, where the new numerical parameter $\varsigma_{I}>0$ that governs the magnitude of energy loss during the interaction could in general differ from the propagation parameter $\zeta$. The presence of such losses affects the relevant threshold equations, for the reaction to occur, which stem from kinematics, in such a way that the stringent limits on $\delta$~(or $E_{\textrm{LV}}$ for $\delta_{(\overline{\nu})}>0$) from the instability of putative superluminal neutrino from IceCube experiments are not valid~(see below for details). 

We have computed both $E_{\textrm{th}}$, the energy thresholds, for the pair emission and splitting by an energetic superluminal antineutrino in D-particle backgrounds: 
\begin{align}
\label{eq:7}
E_{\textrm{th}}({\overline{\nu}\rightarrow\overline{\nu}e^{+}e^{-}})&\simeq\biggl[4m_{e}^{2}\frac{M_{s}}{g_{s}(\zeta^{2}-4\varsigma_{I})}\biggr]^{1/3},\\
\label{eq:8}
E_{\textrm{th}}({\overline{\nu}\rightarrow\overline{\nu}\nu\overline{\nu}})&\simeq\biggl[9m_{\nu}^{2}\frac{M_{s}}{g_{s}(\zeta^{2}-2\varsigma_{I})}\biggr]^{1/3},
\end{align}
where $m_{e}$ is the electron mass.~(Note that Lorentz violation is absent for electrons~(positrons) in D-foam scenarios, for reasons of charge conservation~\cite{Ellis:2003sd,Ellis:2003ua,Ellis:2003if}, and this is justified by the constraints~\cite{Li:2022ugz} on any departure from Lorentz-invariant electron dispersion relations.) These expressions follow from simple calculations of the threshold equations by taking into account the energy violation $\delta E_{D}$. The fact that for pair emission the incoming energy is almost taken by the $e^{+}e^{-}$ pair is used while for splitting we assume that the energy is carried off roughly equally by the 3 outgoing particles. For $\overline{\nu}\rightarrow\overline{\nu}e^{+}e^{-}$, it is thus kinematically allowed when the initial antineutrino energy 
\begin{equation}
\label{eq:9}
E_{\overline{\nu}}>\frac{2m_{e}}{\sqrt{\D}},\quad\D=\Bigl(1-\frac{4\varsigma_{I}}{\zeta^{2}}\Bigr)\delta_{\overline{\nu}},
\end{equation}
or equivalently, $E_{\overline{\nu}}\geq E_{\textrm{th}}\equiv (4m_{e}^{2}{\cal E}_{*})^{1/3}$, here a new energy scale ${\cal E}_{*}$ is defined as
\begin{equation}
\label{eq:10}
{\cal E}_{*}({\overline{\nu}\rightarrow\overline{\nu}e^{+}e^{-}})=\frac{\Elv}{1-4\varsigma_{I}\Elv/M_{D}},
\end{equation}
where $\delta_{\overline{\nu}}=\zeta^{2}E_{\textrm{th}}/M_{D}=E_{\textrm{th}}/\Elv$ at threshold is used. The above threshold is finite only if $4\varsigma_{I}<\zeta^{2}$, otherwise, it would be either infinite~($\zeta^{2}=4\varsigma_{I}$) or imaginary~($\zeta^{2}<4\varsigma_{I}$), implying that pair emission is forbidden in a vacuum. Even in the case of $\zeta^{2}>4\varsigma_{I}$, the process could be effectively inhibited if $\varsigma_{I}\approx (\zeta/2)^{2}$, in which case the threshold would be pushed to a very high energy scale of, for instance order PeV, so that PeV antineutrinos will not be depleted. We are then able to observe superluminal events of such energy at IceCube.~(Lacking, at present, a complete theory of matter-D-foam interactions, we do not discuss the emission rate above threshold but nonetheless up to first order in $1/M_{D}\sim\Ep^{-1}$ it should agree with relativistic field theoretic result.) Thus, the tight bounds thereof, e.g.,~\cite{Borriello:2013ala,Stecker:2013jfa,Diaz:2013wia,Wang:2020tej} are actually imposed on $\D$ or on the scale ${\cal E}_{*}$ in D-foam situations, though can be evaded if ${\cal E}_{*}\leq 0$ or easily satisfied in case of ${\cal E}_{*}>0$ by proper assignment for
the values of $\zeta$ and $\varsigma_{I}$, but \textit{do not} constrain the \textit{actual} superluminal velocity $\delta$ and the corresponding scale $\Elv$. 

Taking into account neutrino splittings exposes more interesting features of the model. We find that:~(a) the splitting channel opens when $\zeta>\sqrt{2\varsigma_{I}}$;~(b) if $\sqrt{2\varsigma_{I}}<\zeta\leq 2\sqrt{\varsigma_{I}}$ the $\overline{\nu}$-decay channel through emitting $e^{+}e^{-}$ pairs is not allowed, although $\overline{\nu}$-splitting can happen;~(c) otherwise, antineutrinos are stable in D-foam backgrounds and never decay. In that case, there is also no need to fine-tune the model parameter to be in accordance with the high-energy neutrino observations. 

Hence, superluminal antineutrino~(like some events suggested by~\cite{Huang:2018ham,Huang:2019etr,Huang:2022xto} from IceCube) can survive from instability in the context of D-foam model, while otherwise would rapidly lose energy during propagation. In this respect, the detection~\cite{IceCube:2013cdg} of event \#35 of energy 2~PeV, an ``advance'' event according to~\cite{Huang:2018ham} and hence probably a superluminal antineutrino event, implies a limit of $(\zeta^{2}-4\varsigma_{I})\lesssim 8.4\times 10^{-8}$ for $\zeta>2\sqrt{\varsigma_{I}}$.\footnote{Some studies suppose that \#35 might be emitted by a blazar~\cite{Kadler:2016ygj,Wang:2016lne}, so the IceCube \#20 of $\sim 1.1$~PeV~\cite{IceCube:2013cdg}, also an ``advance'' event~\cite{Huang:2018ham}, implies a more reliable limit, $\zeta^{2}-4\varsigma_{I}\lesssim 4\times 10^{-7}$.} This arises from the consideration that the threshold for the most efficient channel $\overline{\nu}\rightarrow\overline{\nu}e^{+}e^{-}$ should be of $\O(\textrm{PeV})$ for observing events of such energy. Higher energy events that may be detected by experiments in the future are likely to yield even stronger constraints on the difference between $\zeta^{2}$ and $4\varsigma_{I}$~(or, $2\varsigma_{I}$). In cases that exceedingly small values of, e.g., $(\zeta^{2}-4\varsigma_{I})$, are required to reconcile with observations, a more natural solution might have to be invoked, namely, no splitting or pair emission should ever happen for antineutrinos despite being slightly superluminal. This can be naturally realized by arranging $\varsigma_{I}\geq\zeta^{2}/2$, as explained. 

\section{Discussion and remarks}

We have mentioned that the D-particle foam has no effect on charged excitations such as electrons due to gauge invariance~\cite{Ellis:2003sd,Ellis:2003ua,Ellis:2003if}, leaving only neutral particles particularly neutrinos being affected by the foam. This difference indicates an induced breaking of the $\textrm{SU}(2)$ symmetry between neutrinos and charged leptons. Hence, Lorentz violation in neutrinos \textit{does not} imply similar sensitivities for LV in charged leptons, or vice versa, so the tight limits cast by, e.g., the absence of Cherenkov effect by PeV electrons in the Crab Nebula from recent LHAASO experiment~\cite{Li:2022ugz}, impose \textit{no} constraint on $\zeta$, unlike the case of, e.g.,~\cite{Crivellin:2020oov}, in which the LV theory is manifestly $\textrm{SU}(2)_{L}$-invariant. The transparency of the foam to those charged probes also provides a microscopic explanation for the fact that for electrons~(or positrons) no deviations from Lorentz invariance have been observed with a precision~\cite{Li:2022ugz}: 
\begin{equation}
\label{eq:10}
\lvert\delta_{e}\rvert\sim 10^{-20},
\end{equation}
in the ultrarelativistic limit where $m_{e}$ can be ignored. Indeed, $\delta_{e}\approx 0$ is expected by the stringy foam model except for the mass effect. The difficulty encountered by~\cite{Crivellin:2020oov} does not arise here. 

We emphasize again that studies of~\cite{Amelino-Camelia:2016fuh,Amelino-Camelia:2016ohi,Huang:2018ham,Huang:2019etr,Huang:2022xto} cannot be said to have yet provided any evidence for violations of CPT or Lorentz invariance without enough confidence on the strategy of analysis. This means that, even tentatively assuming the LV model is correct, among the analyses therein it is likely that there exist probabilities of obtaining such results accidentally, without any intervening QG effect; but nonetheless, as we have argued, they might in some sense serve as a potential support for the theory discussed here, if such associations are supported by more studies. 

Note, moreover, that superluminal (anti)neutrinos as a result of CPT breaking may \textit{not} be referred to as tachyons with imaginary masses, since as we have seen, they do have physical masses that are real, but instead possess Lorentz-violating dispersion relations. 

Certainly the inhibition of the anomalous decay process by these neutrinos is realized in the present context. That may not be unnatural since Lorentz invariance is restored {\it on average} by the zero mean of $\lambda$ despite the presence of recoil $u\sim g_{s}\lambda E/M_{s}$. We conjecture that the model may be characterized by a reduced Lorentz symmetry but preserve covariance, and thus might fall into the covariant scheme of Lorentz violation in which the challenge brought about by superluminality could also be resolved, as was discussed elsewhere~\cite{Ma:2011jj} for other purposes. 

The conclusion of all this is that, according to the presented string-inspired model of space-time foam with bulk D-brane defects, a CPT-violating propagation for the neutrino species may be permitted through the mechanism as contemplated here. We may expect the effects of the CPT violation, and thus of the possible violation of causality to be small in this scenario, although in the case of superluminal velocity the causality issue deserves attention and may be postponed to future study. Otherwise, we consider it very gratifying that such a stringy model is supported by the phenomenological conjecture of neutrino speed variation~\cite{Huang:2018ham,Huang:2019etr,Huang:2022xto} with some suppression scales $\gtrsim 10^{17}$~GeV. A careful analysis of additional data from IceCube and other neutrino telescopes is anticipated to~(dis)prove the speculation and, in particular, to probe or at least constrain the model as well as stringy QG effect. 

\section*{Acknowledgments}

This work is supported by National Natural Science Foundation of China~(Grant No.~12075003). 



\begin{thebibliography}{99}

\bibitem{Amelino-Camelia:1997ieq}
G.~Amelino-Camelia, J.~Ellis, N.~E.~Mavromatos, D.~V.~Nanopoulos, S.~Sarkar, 
\MYhref[journalLinks]{https://doi.org/10.1038/31647}{Nature {393} (1998) 763}.

\bibitem{He:2022gyk}
For a recent review, see, e.g.,
P.~He, B.-Q.~Ma, 
\MYhref[journalLinks]{https://doi.org/10.3390/universe8060323}{Universe {8} (2022) 323}.

\bibitem{Jacob:2006gn}
U.~Jacob, T.~Piran, 
\MYhref[journalLinks]{https://doi.org/10.1038/nphys506}{Nature Phys. {3} (2007) 87}.

\bibitem{IceCube:2013cdg}
M.~G.~Aartsen {et al}.~[IceCube Collaboration], 
\MYhref[journalLinks]{https://doi.org/10.1103/PhysRevLett.111.021103}{Phys. Rev. Lett. {111} (2013) 021103};
\MYhref[journalLinks]{https://doi.org/10.1126/science.1242856}{Science {342} (2013) 1242856};
\MYhref[journalLinks]{https://doi.org/10.1103/PhysRevLett.113.101101}{Phys. Rev. Lett. {113} (2014) 101101}.

\bibitem{IceCube:2016uab}
M.~G.~Aartsen {et al}.~[IceCube Collaboration], 
\MYhref[journalLinks]{https://doi.org/10.1103/PhysRevLett.117.241101}{Phys. Rev. Lett. {117} (2016) 241101};
\MYhref[journalLinks]{https://doi.org/10.1103/PhysRevLett.119.259902}{{119} (2017) 259902(E)}.

\bibitem{IceCube:2014jkq}
M.~G.~Aartsen {et al}.~[IceCube Collaboration], 
\MYhref[journalLinks]{https://doi.org/10.1088/2041-8205/805/1/L5}{Astrophys. J. Lett. {805} (2015)  L5};
\MYhref[journalLinks]{https://doi.org/10.3847/0004-637X/824/2/115}{Astrophys. J. {824} (2016) 115};
\MYhref[journalLinks]{https://doi.org/10.3847/1538-4357/aa7569}{{843} (2017) 112}.

\bibitem{Amelino-Camelia:2016fuh}
G.~Amelino-Camelia, L.~Barcaroli, G.~D'Amico, N.~Loret, G.~Rosati, 
\MYhref[journalLinks]{https://doi.org/10.1016/j.physletb.2016.07.075}{Phys. Lett. B {761} (2016) 318}.

\bibitem{Amelino-Camelia:2016ohi}
G.~Amelino-Camelia, G.~D'Amico, G.~Rosati, N.~Loret, 
\MYhref[journalLinks]{https://doi.org/10.1038/s41550-017-0139}{Nature Astron. {1} (2017) 0139}.

\bibitem{Huang:2018ham}
Y.~Huang, B.-Q.~Ma, 
\MYhref[journalLinks]{https://doi.org/10.1038/s42005-018-0061-0}{Commun. Phys. {1} (2018) 62}.

\bibitem{Huang:2019etr}
Y.~Huang, H.~Li, B.-Q.~Ma, 
\MYhref[journalLinks]{https://doi.org/10.1103/PhysRevD.99.123018}{Phys. Rev. D {99} (2019) 123018}.

\bibitem{Huang:2022xto}
Y.~Huang, B.-Q.~Ma, 
\MYhref[journalLinks]{https://doi.org/10.1016/j.fmre.2022.05.022}{Fundamental Research 
(2022) doi:10.1016/j.fmre.2022.05.022}, 
\MYhref[eprintLinks]{https://arxiv.org/abs/2211.00231}{{arXiv:2211.00231}}.


\bibitem{Ellis:2008fc}
J.~Ellis, N.~Harries, A.~Meregaglia, A.~Rubbia, A.~S.~Sakharov, 
\MYhref[journalLinks]{https://doi.org/10.1103/PhysRevD.78.033013}{Phys. Rev. D {78} (2008) 033013}.

\bibitem{Wang:2016lne}
Z.-Y.~Wang, R.-Y.~Liu, X.-Y.~Wang, 
\MYhref[journalLinks]{https://doi.org/10.1103/PhysRevLett.116.151101}{Phys. Rev. Lett. {116} (2016) 151101}.

\bibitem{Kadler:2016ygj}
M.~Kadler {et al}. 
\MYhref[journalLinks]{https://doi.org/10.1038/NPHYS3715}{Nature Phys. {12} (2016) 807}.

\bibitem{Ellis:2018ogq}
J.~Ellis, N.~E.~Mavromatos, A.~S.~Sakharov, E.~K.~Sarkisyan-Grinbaum, 
\MYhref[journalLinks]{https://doi.org/10.1016/j.physletb.2018.11.062}{Phys. Lett. B {789} (2019) 352}.

\bibitem{Laha:2018hsh}
R.~Laha, 
\MYhref[journalLinks]{https://doi.org/10.1103/PhysRevD.100.103002}{Phys. Rev. D {100} (2019) 103002}.

\bibitem{Wei:2018ajw}
J.-J.~Wei {et al}., 
\MYhref[journalLinks]{https://doi.org/10.1016/j.jheap.2019.01.002}{JHEAp {22} (2019) 1}.

\bibitem{IceCube:2018dnn}
The IceCube Collaboration, Fermi-LAT, MAGIC {et al}., 
\MYhref[journalLinks]{https://doi.org/10.1126/science.aat1378}{Science {361} (2018) eaat1378};
M.~G.~Aartsen {et al}~[IceCube Collaboration], 
\MYhref[journalLinks]{https://doi.org/10.1126/science.aat2890}{Science {361} (2018) 147}.

\bibitem{Zhang:2018otj}
X.~Zhang, B.-Q.~Ma, 
\MYhref[journalLinks]{https://doi.org/10.1103/PhysRevD.99.043013}{Phys. Rev. D {99} (2019) 043013}.

\bibitem{Ellis:1992eh}
J.~Ellis, N.~E.~Mavromatos, D.~V.~Nanopoulos, 
\MYhref[journalLinks]{https://doi.org/10.1016/0370-2693(92)91478-R}{Phys. Lett. B {293} (1992) 37}.

\bibitem{Ellis:1999jf}
J.~Ellis, N.~E.~Mavromatos, D.~V.~Nanopoulos, 
\MYhref[journalLinks]{https://doi.org/10.1103/PhysRevD.61.027503}{Phys. Rev. D {61} (1999) 027503}.

\bibitem{Ellis:2000sf}
J.~Ellis, N.~E.~Mavromatos, D.~V.~Nanopoulos, 
\MYhref[journalLinks]{https://doi.org/10.1103/PhysRevD.63.124025}{Phys. Rev. D {63} (2001) 124025}.

\bibitem{Ellis:2004ay}
J.~Ellis, N.~E.~Mavromatos, M.~Westmuckett, 
\MYhref[journalLinks]{https://doi.org/10.1103/PhysRevD.70.044036}{Phys. Rev. D {70} (2004) 044036}.

\bibitem{Mavromatos:2005bu}
N.~E.~Mavromatos, S.~Sarkar, 
\MYhref[journalLinks]{https://doi.org/10.1103/PhysRevD.72.065016}{Phys. Rev. D {72} (2005) 065016}.

\bibitem{Ellis:2008gg}
J.~Ellis, N.~E.~Mavromatos, D.~V.~Nanopoulos, 
\MYhref[journalLinks]{https://doi.org/10.1016/j.physletb.2008.06.029}{Phys. Lett. B {665} (2008) 412}.

\bibitem{Li:2009tt}
T.~Li, N.~E.~Mavromatos, D.~V.~Nanopoulos, D.~Xie, 
\MYhref[journalLinks]{https://doi.org/10.1016/j.physletb.2009.07.062}{Phys. Lett. B {679} (2009) 407}.

\bibitem{Mavromatos:2012ii}
N.~E.~Mavromatos, S.~Sarkar, 
\MYhref[journalLinks]{https://doi.org/10.1140/epjc/s10052-013-2359-0}{Eur. Phys. J. C {73} (2013) 2359}.

\bibitem{Li:2021gah}
C.~Li, B.-Q.~Ma, 
\MYhref[journalLinks]{https://doi.org/10.1016/j.physletb.2021.136443}{Phys. Lett. B {819} (2021) 136443}.

\bibitem{Crivellin:2020oov}
A.~Crivellin, F.~Kirk, M.~Schreck, 
\MYhref[journalLinks]{https://doi.org/10.1007/JHEP04(2021)082}{JHEP {04} (2021) 082}.

\bibitem{IceCube:2017qyp}
See, e.g., M.~G.~Aartsen {et al.} [IceCube Collaboration], 
\MYhref[journalLinks]{https://doi.org/10.1038/s41567-018-0172-2}{Nature Phys. {14} (2018) 961}.

\bibitem{Cohen:2011hx}
A.~G.~Cohen, S.~L.~Glashow, 
\MYhref[journalLinks]{https://doi.org/10.1103/PhysRevLett.107.181803}{Phys. Rev. Lett.  {107} (2011) 181803}.

\bibitem{OPERA:2011ijq}
T.~Adam {et al}.~[OPERA Collaboration], 
\MYhref[journalLinks]{https://doi.org/10.1007/JHEP10(2012)093}{JHEP {10} (2012) 093}.

\bibitem{Antonello:2012be}
See, e.g., M.~Antonello {et al}.~[ICARUS Collaboration], 
\MYhref[journalLinks]{https://doi.org/10.1007/JHEP11(2012)049}{JHEP {11} (2012) 049};
P.~Adamson {et al}.~[MINOS Collaboration, NIST, and USNO], 
\MYhref[journalLinks]{https://doi.org/10.1103/PhysRevD.92.052005}{Phys. Rev. D {92} (2015) 052005};
K.~Abe {et al}.~[T2K Collaboration], 
\MYhref[journalLinks]{https://doi.org/10.1103/PhysRevD.93.012006}{Phys. Rev. D {93} (2016) 012006}.

\bibitem{Ma:2012zd}
B.-Q.~Ma, 
\MYhref[journalLinks]{https://doi.org/10.1142/S2010194512005910}{Int. J. Mod. Phys. Conf. Ser. {10} (2012) 195}.

\bibitem{Borriello:2013ala}
E.~Borriello, S.~Chakraborty, A.~Mirizzi, P.~D.~Serpico, 
\MYhref[journalLinks]{https://doi.org/10.1103/PhysRevD.87.116009}{Phys. Rev. D {87} (2013) 116009}.

\bibitem{Stecker:2013jfa}
F.~W.~Stecker, 
\MYhref[journalLinks]{https://doi.org/10.1016/j.astropartphys.2014.02.007}{Astropart. Phys. {56} (2014) 16}.

\bibitem{Diaz:2013wia}
J.~S.~Diaz, V.~A.~Kosteleck\'y, M.~Mewes, 
\MYhref[journalLinks]{https://doi.org/10.1103/PhysRevD.89.043005}{Phys. Rev. D {89} (2014) 043005}.

\bibitem{Wang:2020tej}
K.~Wang, S.-Q.~Xi, L.~Shao, R.-Y.~Liu, Z.~Li, Z.-K.~Zhang, 
\MYhref[journalLinks]{https://doi.org/10.1103/PhysRevD.102.063027}{Phys. Rev. D {102} (2020) 063027}.

\bibitem{Ellis:2003sd}
J.~R.~Ellis, N.~E.~Mavromatos, A.~S.~Sakharov, 
\MYhref[journalLinks]{https://doi.org/10.1016/j.astropartphys.2003.12.001}{Astropart. Phys. {20} (2004) 669}.

\bibitem{Ellis:2003ua}
J.~Ellis, N.~E.~Mavromatos, D.~V.~Nanopoulos, A.~S.~Sakharov, 
\MYhref[journalLinks]{https://doi.org/10.1038/nature02481}{Nature {428} (2004) 386}.

\bibitem{Ellis:2003if}
J.~Ellis, N.~E.~Mavromatos, D.~V.~Nanopoulos, A.~S.~Sakharov, 
\MYhref[journalLinks]{https://doi.org/10.1142/S0217751X04019780}{Int. J. Mod. Phys. A {19} (2004) 4413}.

\bibitem{Li:2022ugz}
C.~Li, B.-Q.~Ma, 
\MYhref[journalLinks]{https://doi.org/10.1016/j.physletb.2022.137034}{Phys. Lett. B {829} (2022) 137034}.

\bibitem{Ma:2011jj}
B.-Q.~Ma, 
\MYhref[journalLinks]{https://doi.org/10.1142/S0217732312300054}{Mod. Phys. Lett. A {27} (2012) 1230005}.

\end{thebibliography}
\end{document}